# Potential Cluster Calculations for the Astrophysical *S*-factor and Reaction Rate of the Radiative $^3$He($^4$He,γ)$^7$Be Capture

S B Dubovichenko,[1,2][*] N A Burkova,[2] and A V Dzhazairov-Kakhramanov[1][*]

[1] Fesenkov Astrophysical Institute "NCSRT" NSA RK, 050020, Almaty, Kazakhstan

[2] Al-Farabi Kazakh National University, 050000, Almaty, Kazakhstan

*dubovichenko@mail.ru*
*natali.burkova@gmail.com*
*albert-j@yandex.ru*

**Abstract:** In the frame of a modified potential cluster model based on the classification of orbital states according to Young diagrams and revised interaction potential parameters for the bound states of $^7$Be in the $^3$He$^4$He cluster model with forbidden states, the astrophysical *S*-factor for the radiative capture reaction $^3$He($^4$He,γ)$^7$Be has been calculated from 10 keV. The result obtained, *S*(23 keV) = 0.561 keV·b, reproduces the latest experimental data at 23 keV. The calculated and parametrized reaction rate is compared with some known results in the range of temperatures from 0.05 to 5$T_9$.

**Keywords:** astrophysical energies; radiative capture; total cross-section; reaction rate; potential cluster model; forbidden states; Young diagrams.

## 1. Introduction

Radiative $^3$He($^4$He,γ)$^7$Be capture at ultralow energies is of apparent interest for nuclear astrophysics as a part of the proton-proton (pp) fusion cycle. The pp-cycle may be closed by the $^3$He + $^3$He → $^4$He + 2p process [1] or by the $^3$He + $^4$He → $^7$Be + γ reaction promoted by $^4$He accumulated in the pre-stellar stage (see, for example, [2]). The role of the radiative $^3$He($^4$He,γ)$^7$Be capture in pre-stellar nucleosynthesis, when the temperature decreased to 0.3 $T_9$ ($T_9$ = 10$^9$ K) after the Big Bang, is now under discussion [3].

The $^7$Be production in the process $^3$He($^4$He,γ)$^7$Be is used directly for estimating the formation of $^7$Li as a result of $\beta$-decay. Data on this process are used for the calculation of the lithium isotopes ratio $^6$Li/$^7$Li produced in the Big Bang. The recent data on the $^6$Li/$^7$Li isotopic ratio obtained within the framework of the LUNA collaboration and a detailed discussion of the astrophysical aspects of this problem are reported in [4].

The experimental status of the $^3$He($^4$He,γ)$^7$Be reaction is critically reviewed, and the available theoretical descriptions of the production and destruction of $^7$Be at astrophysical energies are discussed in [5] and [6].

The new measurements for the astrophysical *S*-factor performed at the lowest energy of 23 keV are of special interest for the $^3$He($^4$He,γ)$^7$Be reaction [7]. That is why we are returning to this treatment as this is a possible way to check the predictive abilities of our approach [8] and proceed with the corresponding calculations for the reaction rate.

Basing on the two-body potential cluster model (PCM), we succeed in describing the total cross-sections and astrophysical *S*-factors for the radiative capture of more than 30 reactions [9]. The calculations of these reactions are performed on the basis of a modified version of the PCM (MPCM) with forbidden states (FSs) [10] and the classification of orbital states according to Young diagrams. The completely certain success of the MPCM in describing the total cross-sections of processes of this type can be explained by the fact that the potentials of cluster–cluster interaction in the continuous spectrum are built not only on the basis of known elastic scattering phase shifts or the structure of spectra of the resonance levels of the final nucleus but for a discrete spectrum on the basis of a description of the main characteristics of the bound states (BSs). These potentials are also based on the classification of orbital states according to Young diagrams [9], which makes it possible to determine the presence and number of FSs in each partial wave and hence the number of inner nodes of the relative motion radial wave function (WF) [11]. As a result, each partial potential depends not only on the standard quantum numbers set *JLS* but also on the Young diagrams {*f*} [12].

---

[*] Corresponding author

## 2. Results and Discussion

### 2.1. Interaction Potentials

As was shown, for example in [13], the orbital states in the $^3$He$^4$He cluster system of $^7$Be are pure according to Young diagrams. Therefore, the nuclear partial potentials of $^3$He$^4$He interactions are of the form

$$V_{JLS}(r) = V_0(JLS)\exp(-\alpha r^2) + V_{coul}(r)$$

with parameters obtained on the basis of elastic scattering phase shifts, and, depending on the quantum numbers, $JLS$ can be used directly to consider the characteristics of the BS of $^7$Be [9,10,14]. For the Coulomb potential $V_{coul}(r)$ the usual spherical shape was used in [15].

$$V_{coul}(r) = \begin{cases} \dfrac{Z_1 Z_2}{r} & r > R_c \\ Z_1 Z_2 \left(3 - \dfrac{r^2}{R_c^2}\right) \Big/ 2R_c & r < R_c \end{cases}.$$

Earlier, in [8], we refined the main calculated characteristics of the bound states of $^7$Be in the $^3$He$^4$He channel. For this purpose, the potential parameters of the bound $P$-states, given in Table 1, were improved, and now the calculated energy levels completely coincide with the experimental values [16]. The potential describes the elastic scattering $S$-phase shift [9] from work [17] well, since the transitions from the $S$-waves to the ground state (GS) and to the first excited state (FES) of $^7$Be make the predominant contribution to the astrophysical $S$-factor. Such potentials in the $S$-wave have two forbidden BSs that correspond to the forbidden Young diagrams {7} and {52}. In the $P$-wave, diagram {61} is forbidden along with the bound allowed state (AS) with the Young diagram {43}. In the $D$-wave, there is an FS with diagram {52} [12,13,18], and there is no AS.

Table 1. Potential parameters for GS and FES. $\alpha = 0.15747$ fm$^{-2}$ and $R_c = 3.095$ fm. Energy spectra and charge radius of $^7$Be. The number of FSs is indicated.

| $L_J$ | $V_0$, MeV | $E$, MeV | FSs | $r_{rms}$, fm | $C_W$ |
|---|---|---|---|---|---|
| $^2S_{1/2}$ | -67.5 | – | 2 | – | --- |
| $^2P_{3/2}$ | –83.589554 | –1.586600 | 1 | 2.64 | 5.04(2) |
| $^2P_{1/2}$ | –81.815179 | –1.160820 | 1 | 2.69 | 4.64(2) |
| $^2D_{5/2}$ | -69.0 | – | 1 | – | |
| $^2D_{3/2}$ | -66.0 | – | 1 | – | |

The energies of the bound levels of the considered nuclei in the given potentials were calculated by the finite-difference method [19] with accuracy not worse than $10^{-6}$ MeV. The depths of the potentials in Table 1 were determined based on the description of charge radii and asymptotic constants [8]. Table 1 also shows the results of calculation of the charge radii. To find the nuclear charge radius, we used the cluster radii given in [20].

To control the stability of the asymptotics of the wave function for the ground and first excited bound states at large distances, we used the dimensionless asymptotic constant (AC) $C_W$ of the form [21]

$$\chi_L(r) = \sqrt{2k_0}\, C_W W_{-\eta L+1/2}(2k_0 r),$$

where $\chi_L(r)$ is the numerical wave function of the bound state obtained from the solution of the radial Schrödinger equation and normalized to unity, $W_{-\eta L+1/2}$ is the Whittaker function, which determines the asymptotic behavior of the WF and is the solution of the same equation without nuclear potential; that is, at large distances $r$, $k_0$ is the wave number corresponding to the channel binding energy; $\eta$ is the Coulomb parameter, and $L$ is the orbital angular momentum. Our dimensionless AC $C_W$ is connected with the ANC (asymptotic normalization coefficient) by the formula $A_{NC} = \sqrt{S_f}\,\sqrt{2k_0}\,C_W$, where $S_f$ is the spectroscopic factor! Both $C_W$ and $A_{NC}$ are used in different experimental and theoretical papers. We give all the corresponding references in our work.

As a result, the AC value of 5.03(2) was obtained for the GS, and 4.64(2) was found for the FES. The above error is determined by averaging the resultant calculation of the constants over the interval of 6–16 fm. For root-mean-square (*rms*) charge radii has been obtained: 2.69 fm for the FES and 2.64 fm for the GS.

Based on an analysis of various experimental data for the ground state of $^7$Be in the $^3$He$^4$He channel, the AC of 4.78(7) fm$^{-1/2}$ was suggested in [22]. If the value of the spectroscopic factor is taken as $S_f = 1.5$, as given on average in [23], from $A_{NC} = \sqrt{S_f} C$ one can obtain a value of 3.90(6) fm$^{-1/2}$ for the AC $C$. Because $C = \sqrt{2k_0} C_W$, we reduced it to the dimensionless value $C_W = 4.58(7)$ at $\sqrt{2k_0} = 0.852$ fm$^{-1/2}$, which is somewhat less than our calculated one. For the first excited state, a value of 4.24(6) fm$^{-1/2}$ is given, and at $\sqrt{2k_0} = 0.788$ fm$^{-1/2}$ we obtained $C_W = 4.39(7)$ at $S_f = 1.5$, which is also less. However, in [23], possible values of 1 to 2 are given for the spectroscopic factor $S_f$. Therefore, if we use $S_f = 1.23$, we obtain the value $C_W = 5.05$, which is in good agreement with the above given value from Table 1 for the GS and $C_W = 4.85$ for the FES. Certainly, the $S_f$ value for the FES can be slightly different from that for the GS. So, in the paper (Mohr et al., 1993 [24]), the following values of spectroscopic factors $S_1 = 1.174$ and $S_2 = 1.175$ were used, which were taken from (Kurath and Millener, 1975 [25]).

## 2.2. Capture *S*-factor and reaction rate

Furthermore, following the publication of new experimental data, we consider again the astrophysical *S*-factor for the $^3$He($^4$He,$\gamma$)$^7$Be radiative capture at the lowest energies. As before, we use the potential cluster model [9,10] with FSs and GS potentials for $^7$Be refined here (see Table 1) [8]. In calculations, only the *E*1 transitions are taken into account, since the contributions of the *E*2 and *M*1 transitions are two to three orders of magnitude smaller. In treating the system, the *E*1 transition is possible between the ground $P_{3/2}$-state of $^7$Be and the $S_{1/2}$, $D_{3/2}$, and $D_{5/2}$-scattering states and also between the first excited bound $P_{1/2}$-state and the $S_{1/2}$ and $D_{3/2}$ scattering states.

The calculation results of the astrophysical *S*-factor of the $^3$He($^4$He,$\gamma$)$^7$Be radiative capture at the energies from 10 keV are shown in Figure 1 by the blue solid curve. The dotted curve shows the *S*-factor for capture into the GS, while dotted-dashed curve shows the *S*-factor for capture into the FES. The experimental data are taken from [7,26–31]. For comparison, the *ab initio* [32] and the *R*-matrix calculations [33] are given by curves 1 and 2, respectively.

As can be seen in Figure 1, the results of our calculations at 23 keV lie in the region of experimental errors [7]. At the energy of 20 keV, our calculation yields an *S*-factor of 0.570 keV·b, and at 23 keV it is 0.561 keV·b. In our earlier works [8,9], $S(20\,\text{keV}) = 0.593$ keV·b was obtained, which is approximately 4% higher than the present results. The reason for this is that the same potentials were used for the GS and the FES earlier. At that time, this was of no principle importance, since the errors of the *S*-factor in the previously measured energy region from 90 keV and above were 10–20% and data at lower energies were absent.

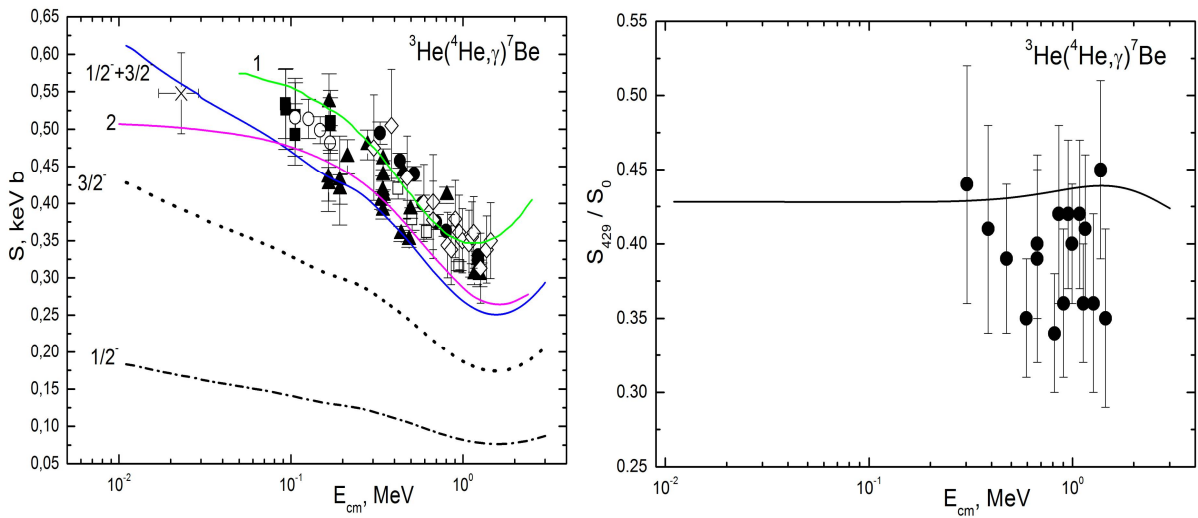

**Figure 1.** Astrophysical *S*-factor of the $^3$He($^4$He,$\gamma$)$^7$Be radiative capture at the energies from 10 keV (left panel). Experiment: solid points – [26]; solid squares – [27]; open points – [28]; open squares – [29]; triangles – [30]; open diamonds – [31]; crosses – [7]. Theory: curve 1 – [32]; curve 2 – [33]; for details of the present calculations with the parameters of Table 1 see the text (blue solid curve, black dotted and dotted-dashed curves). Ratio $S_{429}/S_0$ (right panel) – ratio between *S*-factor of the FES and *S*-factor of the GS: points from [31]; solid curve – present calculations.

The most recent measurements of the S-factor at $23^{+6}_{-5}$ keV [7] lead to the value of 0.548(54) keV·b, which agrees well with our results. A combined average of S(0) = 0.561 ± 0.014 keV·b is found in [7].

For comparison, we present the results of extrapolating the experimental data to zero energy in units of keV·b: 0.54(9) [36], 0.550(12) [34], 0.595(18) [26], 0.560(17) [27], 0.550(17) [28], and 0.567(18) [35]. All these data, instead of the oldest data from [26], within limits of errors, lead to the values that coincide with the new results [7]. In addition, not so long ago, in [23], based on the analysis of various experimental data, the following recommended values were obtained: S(0) = 0.613 keV·b and S(23) = 0.601 keV·b. The latest publication [22] reported the value S(0) = 0.596(17) keV·b, which is slightly higher than the results of [7].

Note that our calculations of the S-factor were performed in [8] in 2010. Only a small refinement has been made here due to the use of the corrected FES potential. New experimental data [7] were published in 2015. In other words, the theoretical results of [8] predicted the behavior of the S-factor at the lowest energies up to 23 keV.

The reaction rate calculated in the $T_9$ range from 0.05 to 5 $T_9$ according to the traditional definition [36] is

$$N_A \langle \sigma v \rangle = 3.7313 \cdot 10^4 \mu^{-1/2} T_9^{-3/2} \int_0^\infty \sigma(E) E \exp(-11.605 E / T_9) dE,$$

where E is taken in MeV, the cross-section σ(E) in μb, the reduced mass μ in amu, and the temperature $T_9$ in $10^9$ K.

The solid curve in Figure 2 (left panel) shows the reaction rate for the total capture, which corresponds to the blue solid curve in Figure 1 (present result). For comparison, the rates from [7], [31], and [33] are given. Note, we chose the latest data, and provided by the analytical parametrizations for the $\langle \sigma v \rangle$ only.

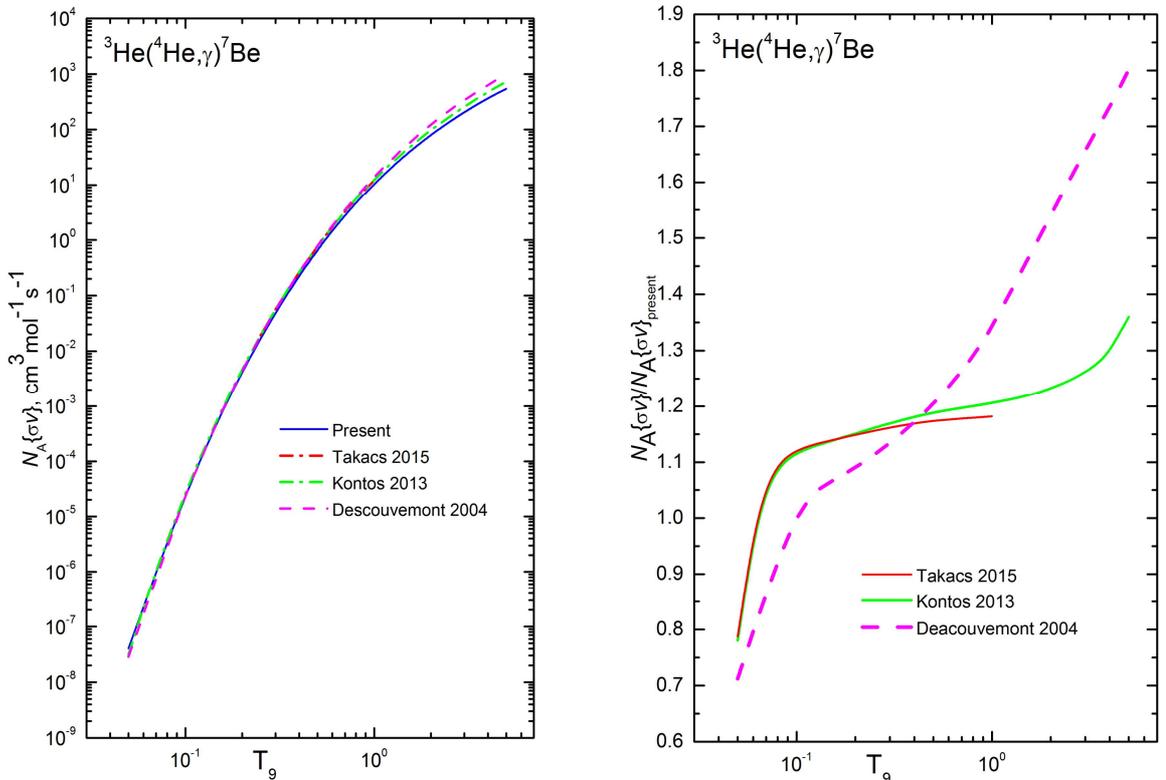

**Figure 2.** Left panel: Reaction rate in the $T_9$ range from 0.05 to 5 $T_9$. Right panel: thermonuclear reaction rate, relative to the present rate, obtained by Kontos *et al.* (green curve, [31]), by Takács *et al.* (red curve, [7]), and by Descouvemont *et al.* (magenta curve, [33]).

To see the real range of deviations in the reaction rates, the corresponding ratios are given in Figure 2. It can be clearly seen that our results are very close to those of Takács *et al.* (red curve, [7]), except at ultra-low energies, where our results are higher. Practically the same ratio is observed for the data by Kontos *et al.* (green curve, [31]) up to ~1 $T_9$, but the higher the value of $T_9$, the more definite the difference that is illuminated. The greatest difference appears in comparison with calculations presented by Descouvemont *et al.* (magenta curve, [33]).

For the total reaction rate in Figure 2, the following analytical parametrization was obtained

$$N_A \langle \sigma v \rangle = 36807.346 / T_9^{2/3} \cdot \exp(-11.354 / T_9^{1/3}) \cdot (1.0 - 15.748 \cdot T_9^{1/3} + 56.148 \cdot T_9^{2/3} + 27.650 \cdot T_9 - 66.643 \cdot T_9^{4/3} + 21.709 \cdot T_9^{5/3}) + 44350.648 / T_9^{3/2} \cdot \exp(-16.383 / T_9),$$

which has an accuracy of $\chi^2$ = 0.91 within 1% theoretical errors.

## 3. Conclusions

The slightly refined variants of calculations for the astrophysical *S*-factor of the $^3$He($^4$He,$\gamma$)$^7$Be reaction are in better agreement with the data available earlier and the latest experimental data. The new lowest measured point $S(23^{+6}_{-5} \text{ keV}) = 0.548 \pm 0,054 \text{ keV} \cdot \text{b}$ (Takács *et al.*, 2015 [7]) is just on the theoretical curve calculated earlier for *S*(E) [8]. So, the predictive reliability of the developing cluster model approach was demonstrated. The new parametrization for the reaction rate is obtained and may be recommended for the astrophysical evaluation of the $^7$Be production. Also we ought to note that at the beginning of this year the paper [37] was published in which the astrophysical *S*-factor of the $^3$He($^4$He,$\gamma$)$^7$Be reaction was obtained, but the reaction rate and its parametrization were not calculated.

We would like to note that a preliminary version of the present results [38] has already found successful application; in particular, the reaction rate analytical parametrization was used in the latest calculations on the solution of the primordial lithium abundance problem of BBN [39].

## Acknowledgments

This work was supported by the Ministry of Education and Science of the Republic of Kazakhstan (Grant No. AP05130104) entitled "Study of Reactions of the Radiative Capture in Stars and in the Controlled Thermonuclear Fusion" through the Fesenkov Astrophysical Institute of the National Center for Space Research and Technology of the Ministry of Defence and Aerospace Industry of the Republic of Kazakhstan (RK).

## References

[1] G Imbriani *Journal of Physics: Conference Series* **312** 042004 (2011)
[2] Ya B Zel'dovich and I D Novikov, *Relativistic Astrophysics, 2: Structure and Evolution of the Universe.* Ed. Gary Steigman (Chicago: Chicago press) (1983)
[3] A Caciolli et al *Eur. Phys. J. A* **39** 179 (2009)
[4] D Trezzi, M Anders, M Aliotta et al *Astroparticle Physics* **89** 57 (2017)
[5] A Di Leva, L Gialanella and F Strieder *Journal of Physics: Conference Series* **665** 012002 (2016)
[6] A S Solovyev and S Yu Igashov *Phys. Rev. C* **96** 064605 (2017)
[7] M P Takács et al *Phys. Rev. D* **91** 123526 (2015)
[8] S B Dubovichenko *Phys. Atom. Nucl.* **73** 1517 (2010)
[9] S B Dubovichenko *Thermonuclear processes in Stars and Universe* (Saarbrücken: Lambert Acad. Publ.) (2015)
[10] S B Dubovichenko, A V Dzhazairov-Kakhramanov *Int. Jour. Mod. Phys. E* **26** 1630009(1-56) (2017)
[11] V G Neudatchin, A A Sakharuk, and S B Dubovichenko *Few-Body Systems* **18** 159 (1995)
[12] V G Neudatchin et al *Phys. Rev. C* **45** 1512 (1992)
[13] V I Kukulin, V G Neudatchin and Yu F Smirnov *Sov. Jour. Part. Nucl.* **10** 1236 (1979)
[14] S B Dubovichenko *Radiative neutron capture and primordial nucleosynthesis of the Universe* (Saarbrücken: Lambert Acad. Publ.) (2016) (in Russian)
[15] P E Hodgson *The Optical model of elastic scattering* (Oxford: Oxford press) (1963)
[16] D R Tilley et al *Nucl. Phys. A* **708** 3 (2002)
[17] A. C Barnard, C M Jones and G C Phillips *Nucl. Phys.* **50** 629 (1964)
[18] S B Dubovichenko *Light nuclei and nuclear astrophysics* (Saarbrücken: Lambert Acad. Publ.) (2013) (in Russian)
[19] S B Dubovichenko *Methods of Calculation of Nuclear Characteristics: Nuclear and Thermonuclear Processes* (Saarbrücken: Lambert Acad. Publ.) (2012) (in Russian)
[20] Centre for Photonuclear Experiments Data; http://cdfe.sinp.msu.ru
[21] G R Plattner and R D Viollier *Nucl. Phys. A* **365** 8 (1981)
[22] R Yarmukhamedov, O R Tojiboev and S V Artemov *Nuovo Cim.* **39 C** 364 (2016)

[23] K I Tursunmakhatov and R. Yarmukhamedov *Phys. Rev. C* **85** 45807 (2012)
[24] P. Mohr et al *Phys. Rev. C* **48** 1420 (1993)
[25] D. Kurath and D.J. Millener *Nucl. Phys. A* **238** 269 (1975)
[26] C R Brune, R W Kavanagh and C Rolf *Phys. Rev. C* **50** 2205 (1994)
[27] F Confortola et al *Phys. Rev. C* **75** 065803 (2007)
[28] G Gyurky et al *Phys. Rev. C* **75** 035805 (2007)
[29] N Singh et al *Phys. Rev. Lett.* **93** 262503 (2004)
[30] J L Osborne et al *Nucl. Phys. A* **419** 15 (1984)
[31] A Kontos et al *Phys. Rev. C* **87** 065804 (2013)
[32] T Neff *Phys. Rev. Lett.* **106** 042502 (2011)
[33] P Descouvemont et al *Atom. Data and Nucl. Data Tables* **88** 203 (2004)
[34] D Bemmerer et al *Phys. Rev. Lett.* **97** 122502 (2006)
[35] H. Costantini et al *Nucl. Phys. A* **814** 144 (2008)
[36] C Angulo et al *Nucl. Phys. A* **656** 3-183 (1999)
[37] E M Tursunov, S A Turakulov and A S Kadyrov *Phys. Rev. C* **97** 035802 (2018)
[38] S B Dubovichenko, N A Burkova, A V Dzhazairov-Kakhramanov arXiv: 1706.05245v1 [nucl-th]
[39] Vinay Singh et al arXiv:1708.05567v1 [nucl-th]